\documentclass[a4paper,11pt,twocolumn]{article}

\usepackage{icphs2023}
\usepackage{metalogo} 
\usepackage{epstopdf}
\usepackage{multirow}
\usepackage{xcolor}
\usepackage{easyReview}

\title{Evolution of Voices in French Audiovisual Media Across Genders and Age in a Diachronic Perspective.}
\author{Albert Rilliard$^{1,2}$, David Doukhan$^3$, R\'emi Uro$^{1,3}$, Simon Devauchelle$^3$}
\organization{$^1$ Universit\'e Paris Saclay, CNRS, LISN, Orsay, France; $^2$ Federal University of Rio de Janeiro, Brazil;  $^3$ French National Institute of Audiovisual, Paris, France}
\email{albert.rilliard@lisn.upsaclay.fr, \{ddoukhan,ruro,sdevauchelle\}@ina.fr}

\begin{document}

\maketitle

\begin{abstract}
We present a diachronic acoustic analysis of the voice of 1023 speakers from French media archives. 
The speakers are spread across 32 categories based on four periods (years 1955/56, 1975/76, 1995/96, 2015/16), four age groups (20-35; 36-50; 51-65, >65), and two genders. 
The fundamental frequency ($F_0$) and the first four formants (F1-4) were estimated. 
Procedures used to ensure the quality of these estimations on heterogeneous data are described. 
From each speaker's $F_0$ distribution, the base-$F_0$ value was calculated to estimate the register. Average vocal tract length was estimated from formant frequencies. 
Base-$F_0$ and vocal tract length were fit by linear mixed models to evaluate how they may have changed across time periods and genders, corrected for age effects. 
Results show an effect of the period with a tendency to lower voices, independently of gender. 
A lowering of pitch is observed with age for female but not male speakers.
\end{abstract}


\section{Introduction}

Vocal characteristics are an important part of identity, with our voices indicating our gender and other social constructs \cite{Sergeant_Welch_2009}. 
Voice obviously changes with age, a lot during our development until adulthood \cite{Fouquet_Pisanski_Mathevon_Reby_2016}, 
but also later with aging voices \cite{Russell_Penny_Pemberton_1995}. 
Social constructs, and the acoustic characteristics linked to them, may change with cultures -- as was described by van Bezooijen for Japanese and Dutch female voice \cite{vanBezooijen_1995}: 
it is thus essential to study these characteristics in varied cultural and language contexts (and it typically fits this ICPhS theme). 
Some studies also claimed that voices may have changed across time (for populations of comparable age); this was claimed for female voices, for example in Australia \cite{Pemberton_McCormack_Russell_1998}.

We present here a first acoustic analysis of a corpus of voices extracted from French broadcast archives in a diachronic perspective, trying to balance the selection of speakers according to their age and gender. 
A total of 1023 speakers were selected, and samples of their vocal production were extracted from the original archive. 
This corpus aims to extract acoustic cues from a large sample, representative of female and male voices presented in French media, to describe them and their possible changes with time and age.
Some studies have described diachronic changes of voices across genders, but these studies are few, may have varying conclusions or population sampling, and address populations from different cultural backgrounds. 
\cite{Pemberton_McCormack_Russell_1998} addresses young adult women's voices, with two points in time -- while \cite{Zou_Wang_He_2012} studied changes in male voices over five time periods, focusing on media anchors: comparatively few speakers were included (four for the first three periods), with possible longitudinal effects.
For French, \cite{Suire_Barkat-Defradas_2020} selected short samples of voices of anonymous individuals from media archives (mean duration of 5 s.), comparing $F_0$ of males and females across 70 years. 
They report complex patterns of $F_0$ changes across time, with possible opposite tendencies for genders after the 1960s, but don't control for age, a factor known to induce changes in voice $F_0$ \cite{Stathopoulos_Huber_Sussman_2011}.
Vocal characteristics are also affected by vocal effort \cite{Berg_Fuchs_Wirkner_Loeffler_Engel_Berger_2017}, which has substantial effects on $F_0$ \cite{Titze_Sundberg_1992} and on formants \cite{Rilliard_2018}.

Long-term acoustic measures give reliable information on voice characteristics. \cite{Lofqvist_Mandersson_1987} showed the long-term average spectrum stabilizes for articulatory changes with about 10 seconds of voiced speech.
\cite{Arantes_Eriksson_2014} compared several measures linked to register and found the base-$F_0$ \cite{Traunmuller_Eriksson_1995} was faster to stabilize (with less than ten s. of voiced speech) than mean or median $F_0$. 
A study \cite{Johnson_2018} linked to the estimation of the vocal tract length (VTL) from formants showed the method proposed in \cite{Lammert_Narayanan_2015} was the most rapidly stable, even on a reduced dataset.
The literature reports that gender is perceived notably through two acoustic cues -- $F_0$ and supraglottal resonances \cite{Pisanski_Rendall_2011,Leung_Oates_Chan_Papp_2021}.

The remaining parts of the paper present estimates of $F_0$ and VTL made on a corpus collected following the semi-automatic strategy proposed by \cite{uro2022semi} to gather speech samples from speakers across 60 years period, found in TV and Radio archives of the French National Institute of Audiovisual (INA) (speakers belong to different categories of age and gender). 
The acoustic measures related to the speakers' voice pitch are fitted by linear mixed models evaluating potential changes across time for both genders and controlling for age.

\section{Methods}

\subsection{INA's diachronic corpus}

INA's diachronic corpus contains speakers of both genders, spread across four age categories (20-35, 36-50, 51-65, over 65 years old) and four time periods (1955-56, 1975-76, 1995-96, and 2015-16).
With an initial goal of gathering at least 30 individuals for each of these 32 categories of age, gender, and time period, samples of voice for 1023 speakers were collected (see Table \ref{tab:corpusstats}).
Some of these categories being much less present in media (typically women), finding target speakers from all age categories in the earliest periods was challenging. 
It was thus mandatory to increase the targeted period for the 1955-56 and 1975-76 periods by considering the years between 1954-1957 and 1974-1977 (note the additional years only represent a small part of the selected speakers so that these periods will be referred to with the originally targeted years for simplicity).
The collection was done thanks to INA's archivists identifying specific women and men within the four age categories and featuring in programs from the four given targeted time periods. 
Following the method described in \cite{uro2022semi}, the programs featuring target speakers were submitted to a diarization process, and the ID corresponding to each target speaker was then hand-picked. These speakers' samples were then selected to keep samples with minimal noise or background music and remove silences.

These archive extracts were submitted to procedures to discard excerpts with adverse characteristics.
\texttt{LIUM\_SpkDiarization}~\cite{meignier2010lium} was used to reject segments associated with a telephone quality.
Speakers with less than 10 seconds of valid pitch estimates were rejected (the filtering protocol is detailed in section \ref{sec:filt}).
Table \ref{tab:corpusstats} presents the distribution of the speakers. 
Voice samples from 1023 speakers were obtained from 878 radio or TV programs (the voice of 85 persons was collected from more than one program).
The median duration of valid acoustic features per speaker was 125 seconds. 
%
The amount of unique speakers in the 32 categories varies between 15 and 74.

\begin{table}[!ht]
\begin{center}
\begin{tabular}{|c|c c| c c| c c| c c |}
\hline
\rowcolor[gray]{.75}
&  \multicolumn{2}{c|}{20-35} & \multicolumn{2}{c|}{36-50} & \multicolumn{2}{c|}{51-65} & \multicolumn{2}{c|}{\textgreater 65}\\
\rowcolor[gray]{.75}
\multicolumn{1}{|r|}{ } & F & M & F & M & F & M & F & M\\
\hline
1954-57 & 17 & 41 & 22 & 74 & 18 & 50 & 17 & 15 \\
1974-77 & 18 & 17 & 23 & 41 & 28 & 39 & 20 & 26 \\
1995-96 & 32 & 31 & 32 & 46 & 29 & 47 & 29 & 35 \\
2015-16 & 30 & 31 & 30 & 52 & 28 & 48 & 29 & 31 \\
\hline
\end{tabular}
\caption{Number of speakers per time period (rows), age group, and gender in the corpus}\label{tab:corpusstats}
\end{center}
\end{table}

\subsection{Voicing, $F_0$ and Formant estimation}

To estimate robust $F_0$ and formant measurements from widely different materials (due to heterogeneity in recording and archival conditions), the voicing decision, and then $F_0$ and formant estimations, were made by concurrent algorithms.
The \texttt{Spleeter} source separation framework was used to separate voice from other phenomena (music or noise)~\cite{hennequin2020spleeter}. The acoustic features were then estimated twice, from both the original and speech-separated signals.
For voicing and $F_0$ estimation, \cite{vaysse2022} showed good performances of \texttt{Praat} auto-correlation (ac) algorithm; it also shows that these estimations could be improved in noisy conditions by combining \texttt{REAPER}'s voicing estimation with \texttt{FCN-F0}'s $F_0$ estimation (based on neural-network). 
We thus used \texttt{Praat}~\cite{boersma2020praat} for estimating voicing and $F_0$ (ac algorithm with a 65-650 Hz $F_0$ range) and for estimating the first four formants (using the recommended settings for female and male voices: 5 formants for a respective ceiling frequency of 5.5 or 5kHz). 
\texttt{REAPER} algorithm~\cite{talkin2015reaper} was used with two distinct settings (default and Hilbert transform) to obtain two other voicing estimates. 
\texttt{FCN-F0} was used with default pitch range (30-1000 Hz), and Viterbi smoothing to estimate $F_0$ \cite{Ardaillon2019}.

\subsection{Acoustic features filtering}
\label{sec:filt}

The frames detected as voiced by the six estimations 
(\texttt{Praat} and \texttt{REAPER} with two sets of parameters on the raw and separated signals) were kept.
From these frames, those where the four $F_0$ estimations (by \texttt{Praat} and \texttt{FCN-F0} on raw and separated signals) were below a 20\% gross error rate threshold were kept. 
This strategy allowed us to keep 52.2\% of the signal's original frames instead of 61.1\% obtained with \texttt{Praat}'s \texttt{ac} algorithm.
From these frames, the formant estimates made on both signals (raw and source-separated) were compared to keep only frames with four formant estimates, with variations below a 20\% gross-error-rate threshold (keeping about 96\% of the preceding frames) and 50.2\% of the signal's original frames.

\subsection{Long-term  features }



For each speaker, the valid samples (cf. above section) were grouped in chunks of at least 10s; i.e., for a 36s extract, three chunks of 12s were used.
From each of these chunks, two long-term estimates were made: the speaker's base-$F_0$ (defined as the seventh decile of the $F_0$ distribution on the chunk \cite{Traunmuller_Eriksson_1995,Arantes_Eriksson_2014}), and the speaker's vocal tract length (equal to the median of the length estimations made on each frame based on the first four formants, using the equation proposed in \cite{Lammert_Narayanan_2015,Johnson_2018}).
Let's make it clear the estimation of the "vocal tract length" is a way to evaluate the tendency one speaker has to produce higher or lower resonances (formants) as a result of  their articulatory habits; the relation to actual vocal tract length is undoubtedly more complex \cite{titze2015toward}.
For each speaker, these two long-term estimates (base-$F_0$ and vocal tract length: VTL) were estimated for each chunk of voiced samples. The median number of chunks was 14 by speaker, ranging from 1 to 175.

\subsection{Statistical analysis}

The base-$F_0$ and the estimated VTL were fitted with two linear mixed models that took as fixed effects the actual age (in years) of the speaker, the time period of the recording (four levels: 1955-56, 1975-76, 1995-96, 2015-16), and the speaker's gender -- and as random effects the speaker for which the measure was done, and the media program from which the corresponding speaker's chunk was extracted (the program factor was nested in the speaker factor).
Following \cite{Gries_2021}, a maximal model was fit (using R's \texttt{lme4} library \cite{lme4}) that included all the interactions between the fixed factors. This model was then submitted to a simplification procedure to remove non-significant terms and reach a minimal adequate model, one for $F_0$ and one for VTL.
These two models are used here to describe the results of the variation of the base-$F_0$ and VTL across age, gender, and time period.

\section{Results}

\subsection{Base-$F_0$}

\begin{figure}[!ht]
\begin{center}
\includegraphics[width=6cm]{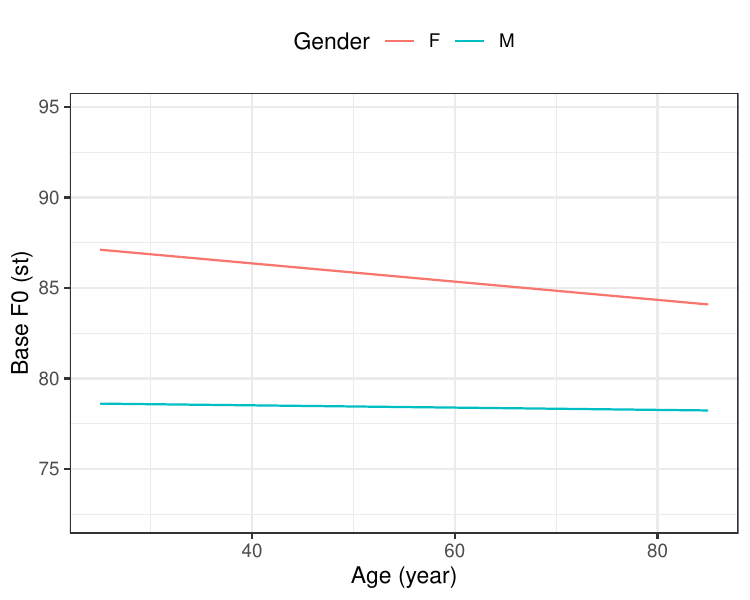}
\caption{Fit of base-$F_0$ (in semitones) from the Speaker Age and Gender.}\label{fig:f0agegender}
\end{center}
\end{figure}

\begin{figure}[!ht]
\begin{center}
\includegraphics[width=6cm]{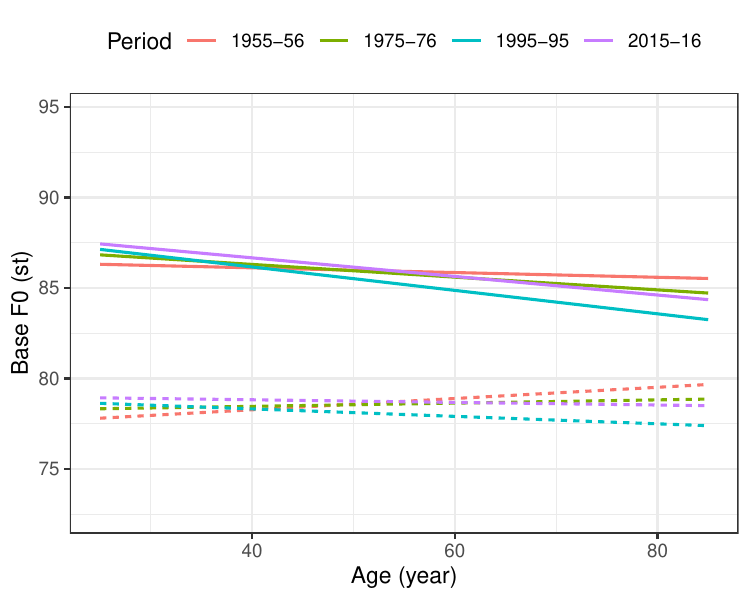}
\caption{Fit of base-$F_0$ (semitones) for speaker Age by Period and separated by Gender (males: dashed lines) to show the relative effect of the Period slopes on voices of each gender.}\label{fig:f0ageperiod}
\end{center}
\end{figure}

The minimal model fitted on base-$F_0$ was based on the three main fixed factors (\texttt{Age}, \texttt{Period} and \texttt{Gender}), with the \texttt{Age:Period} and \texttt{Age:Gender} interactions, plus the original random structure (\texttt{Program} nested in \texttt{Speaker}).
An estimation of the conditional and marginal coefficients of determination (using \cite{mumin}) for this model showed the model explained about 90\% of the variance, with fixed factors responsible for 58\%  of this. 
Figure~\ref{fig:f0agegender} presents the effect of the \texttt{Age:Gender} interaction on base-$F_0$: there is a major difference in $F_0$ across genders (about 9 st), and $F_0$ decreased with age for female voices. 
Figure~\ref{fig:f0ageperiod} presents the \texttt{Age:Period} interaction: while Base-$F_0$ was increasing with Age in the 1950s, it showed a tendency to decrease with age at later periods (1995-96 and 2015-16). 
Note that this second interaction has a much smaller effect size than the \texttt{Age:Gender} one.

\subsection{Vocal tract length}

The minimal model fitted on VTL was based on the three fixed factors without interaction and keeping the original random structure.
The conditional and marginal coefficient of determination for this model showed a much smaller part of the variance was explained by the factors, with the complete model explaining about 38\% and fixed factors 31\%.
The \texttt{Gender} had the largest effect size, with a mean difference of estimated VTL of 1.7 cm between females and males.
\texttt{Period} was the next important factor, with a modest but significant increase in VTL of 0.13 and 0.2 cm for the 1995-96 and 2015-16 periods, compared to 1955-56.
\texttt{Age} showed a significant but small (0.03 cm for ten years) increasing slope of estimated VTL with age.

\section{Discussion \& Conclusion}

This study intends to bring discussions and gather remarks and advice from the community by presenting preliminary results on a complex topic related to the perception of voice quality, and importantly voice pitch, across gender, age, and time periods in France.
As the material used to extract the stimuli could not, for obvious reasons, be of homogeneous quality (recording conditions, signal quality, etc.), there are a series of limitations related to this approach.
A problem is linked with the estimation of vocal tract resonances, as highlighted, e.g., by \cite{titze2015toward}. 
The quality of archive acoustic quality further complicates this: we had to implement a series of checks to assert we didn't process voices with telephone quality or other deterioration linked to, e.g., compression.

The question of the formant estimation algorithm's parameters is essential, as it may lead to varying results.
The data presented here are based on the default parameters recommended by \texttt{Praat} for each gender, but this is a potential bias for estimating VTL in speakers with non-standard characteristics.
We also tested the parameters recommended by \cite{Lammert_Narayanan_2015} (estimate six formants for a 5.5kHz ceiling frequency): this option led to very different results -- and a complex interaction between \texttt{Gender}, \texttt{Age}, and \texttt{Period} with some cases of female speakers having longer estimated VTL than male ones. It was thus not kept here, but questions remain on the estimation of VTL. 
While distinct settings are generally recommended for the analysis of male and female voices (frequency range for $F_0$ estimation, frequency ceiling for formats), the use of these a priori settings for the investigation of gender characteristics is questionable since voices, recording, and archival strategies may have changed over the last 60 years -- and because it may hide some non-standard features.

Conversely, measures of $F_0$ seem much more robust and show trends that confirm the literature, with an effect of age on female voice \cite{Russell_Penny_Pemberton_1995,Yamauchi_Yokonishi_Imagawa_Sakakibara_Nito_Tayama_Yamasoba_2015}, and a clear gender difference for gender representation in French media. 
One finding of this study is linked to a change in the evolution of base-$F_0$ with age across time periods: while it tends to increase in the 1950s for males or stay almost constant for females, steeper slopes were observed in the two later periods (1995-96, and 2015-16). This evolution is not dependent on gender (the interaction was not significant), but the slopes are different for each gender, as shown by figure~\ref{fig:f0ageperiod}. This decreasing tendency raises questions on the use of voice in public displays (varying social use~\cite{boula2012diachronic} or changing health conditions?), with lowered voices for older speakers across time periods.

\section{Acknowledgements}
This work has been funded by the French National Research Agency (project Gender Equality Monitor - ANR-19-CE38-0012). The authors are deeply indebted to INA's archivists La\"etitia Larcher and Anissa-Claire Adgharouamane and  who selected the target speakers and also wished to thank Luc Ardaillon and Jean-Hugues Chenot for their valuable advice.

\bibliographystyle{IEEEtran}
\bibliography{icphs2023}


\end{document}